\documentstyle[aps,prl,epsf,epsfig,twocolumn,multicol]{revtex}  
\newcommand{\be}[1]{\begin{equation}\label{#1}}
\newcommand{\ee}{\end{equation}}     
\newcommand{\bea}{\begin{eqnarray}}
\newcommand{\eea}{\end{eqnarray}} 
\newcommand{\eq}[1]{Eq.\ (\ref{#1})}
\newcommand{\pic}[2]{\epsfxsize #1 cm
\epsffile{sandfig#2.eps}
}
\newcommand{\fig}[1]{Fig.~\ref{#1}}
\begin{document}  
\tighten   
%
\title{ \large\bf Irregular  orbits generate higher harmonics}
\author{Gerd van de Sand$^{1}$ and Jan M. Rost$^{1,2}$} 
\address{$^{1}$ Theoretical Quantum Dynamics, 
Fakult\"at f\"ur Physik, Universit\"at Freiburg,\\
Hermann--Herder--Str.  3, D--79104 Freiburg,
Germany}
\address{$^{2}$ Institute for Advanced Study, Wallotstr. 19, D-14193 Berlin, 
Germany and \\ Max-Planck-Institute for Physics of Complex Systems, 
N\"othnitzer Str. 38, D-01187 Dresden, Germany\\[3mm]
\rm (March 1999)}
\author{
\begin{minipage}{152mm}  
\vspace*{3mm}
The spectrum of higher harmonics in atoms calculated with a 
uniformized semiclassical propagator is presented and it is shown that 
higher harmonic generation is an interference phenomenon which can be 
described semiclassically.  This can be concluded from the good 
agreement with the quantum spectrum.  Moreover, the formation of a 
plateau in the spectrum is specifically due to the interference of 
irregular, time delayed, trajectories with regular orbits without a 
time-delay.  This is proven by the absence of the plateau in an 
artificial semiclassical spectrum generated from a sample of 
trajectories from which the irregular trajectories (only a few 
percent) have been discarded.
\draft
\end{minipage}
}
\maketitle
The generation of higher harmonics (HHG) is an intriguing and 
experimentally well confirmed phenomenon  which results from the 
non-linear response of a microscopic system to a strong laser field 
\cite{HLMM92,HuBa93}.
HHG has been studied in simple but illustrative models numerically and
analytically \cite{KSK92a,Cor93,Lew94,PLKK9k}, for reviews see
\cite{KSK92b,PKK97}. Thereby, two striking features have been identified, 
namely the occurrence of a ``plateau'', i.e. the almost constant 
intensity of the harmonics over a wide range of orders $N$, and the 
sharp ``cutoff'' at a certain maximum order $N_{max}$ of harmonics.
These features have been explained in terms of a simple 
quasiclassical argument \cite{Cor93,Lew94}.

A closer inspection, however, reveals that only the cutoff can be
explained with this argument that involves  a phase 
matching condition  for the semiclassical amplitude imposing  
constraints on the actions of representative classical orbits. In the 
case of an initially bound electron one obtains the 
intuitively appealing picture that the electron  
must return to the nucleus in a certain time correlated with the 
period (frequency) of the laser field to generate higher harmonics 
\cite{Cor93}. 
If the electron has too much energy (which it  would need   to generate 
extremely high harmonics)  it is too fast to fulfill the matching 
condition. Hence, 
the matching condition does explain the cutoff, or more precisely, 
it predicts that the conditions for HHG are unfavorable for $N>N_{max}$.
On the other hand this does not explain the existence and origin of the plateau
for $N<N_{max}$ since   the cut off condition  does not provide a 
reason 
why the probability
for HHG should be  (almost uniformly) high for $N<N_{max}$ as it is found
in  experiments and in numerical simulations.  
Indeed, only in quantum simulations is the plateau  found, classical 
simulations do not yield a plateau. This raises the question whether the 
plateau is due to inherently quantum mechanical effects, such as 
diffraction or tunneling, or if it is a pure interference 
phenomenon that can be explained semiclassically. 

In order to answer this question one must carry out a full semiclassical 
calculation of HHG which has not be done so far. This is probably due 
to considerable technical difficulties since the chaotic 
dynamics of the explicitly time dependent problem renders a 
standard semiclassical treatment (even for one spatial 
degree of freedom) in the framework of the van Vleck 
propagator \cite{vanV22} impossible. However, using a uniformized propagator
following the ideas of Hermann and Kluk \cite{HK84,Kay94a} we have succeeded 
in 
obtaining a converged semiclassical spectrum of HHG. 
Moreover, we are able (i) to prove that HHG is a pure interference 
effect, and (ii) to identify the different types  of trajectories  
which interfere with each other. 

We have performed our calculation with the ``canonical'' model system 
for  the interaction of a strong laser field with a one-electron atom, 
described by the Hamiltonian
\be{hamil}
H = p^{2}/2 + V(x) + E_{0}x\cos\omega t
\ee
where  
$V(x) = -(x^{2}+a^{2})^{-1/2}$ with $a^{2} = 2$ is  the so called  ``soft core''  
potential   (atomic units are used if not stated 
otherwise). With this choice of $a$
 the ground state energy in the potential $V$  corresponds to
that of  hydrogen, $E = -1/2 \, a.u.$.  The other parameters which will be used
are $E_{0} = 0.1 \, a.u.$ and $\omega = 0.0378 \, a.u.$. 
We propagate a  wavepacket $\Psi(x,t)$ according to
\be{prop}
|\Psi(x,t)\rangle =  U(t)|\Psi(x,0)\rangle.
\ee
 The initial wavepacket has its center $x_{0} =E_0/\omega_0^2= 70$ atomic units 
 away from the nucleus (located at $x=0$) and is defined as
 \be{ini}
 \Psi(x,0) = \left(\frac{\gamma^2}{\pi}\right) ^{1/4}
  \exp \left( -\frac{\gamma^2}{2} \Delta_{xx_0}^2 + \frac{i}{\hbar} p_0 \, \Delta_{xx_0} \right)
\ee
\begin{figure}
\pic{8}{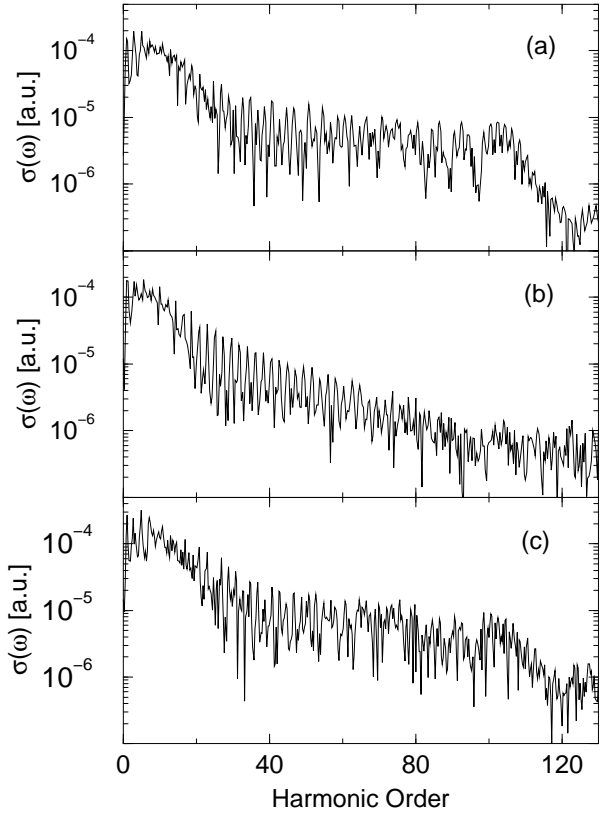}
\caption[]{Quantum (a), classical (b), and semiclassical (c) 
spectrum of higher harmonics according to \eq{spec}.} \label{spectrum}
\end{figure}
\noindent
with $\Delta_{ab} = a-b$,  $\gamma = 0.2236 \, a.u.$ and $p_{0}=0 \, a.u.$. 
Under these conditions of a scattering experiment the cutoff for HHG occurs
at $2U_p = E_0/2\omega^2$, see also \cite{PLKK9k} where 
the same initial conditions have been used 
apart from the  width $\gamma$ which does not occur there.

The Gaussian form of $\psi(x,0)$ allows one to express the 
semiclassically propagated wavefunction in closed form as an integral 
over phase space \cite{Kay94a},
\bea\label{psi}
\Psi(x,t)  &= & \frac{1}{(2 \pi \hbar)} \int \! \! \! \int \! dq \, dp \, \, 
R_\gamma(p_t, q_t)  \, \, \exp \left(\frac{i}{\hbar} S(p_t, q_t) \right) \nonumber \\
&& \exp \left( -\frac{\gamma^2}{2}\Delta_{xq_t}^2 + \frac{i}{\hbar} p_t \Delta_{xq_t}
\right)  \\  & &\exp \left( -\frac{\gamma^2}{4}\Delta_{qx_0}^2 -
\frac{1}{4 \gamma^2}  \Delta_{pp_0}^2 + \frac{i}{2 \hbar}\Delta_{qx_0} (p + p_0) \right) ,
\nonumber
\eea
where 
$S(p_t, q_t) $ is the classical action of a trajectory at 
$t$, and 
\be{monofak}
R_{\gamma}(p_t, q_t) = \left| \frac{1}{2} \left( M_{qq} + 
M_{pp} - i \gamma^2 \hbar \, M_{qp} - \frac{1}{i \gamma^2 \hbar} \,  
 M_{pq} \right) \right| ^{1/2}
\ee
 is composed of all four blocks $M_{ab} = \partial^{2}S/(\partial 
 a\partial b)$ of the monodromy matrix.  
 
\begin{figure}
 \pic{8}{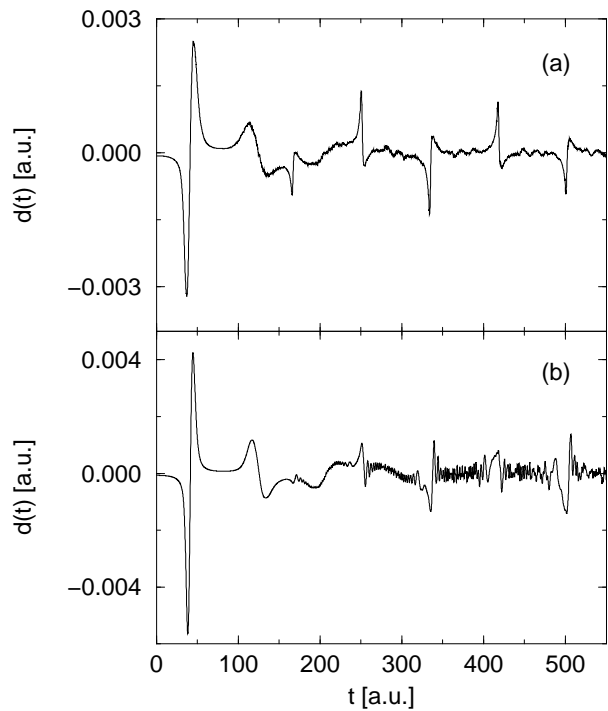}
\caption[]{Classical (a) and  semiclassical (b) dipole acceleration according to 
\eq{dipol}.} \label{dipolpct}
\end{figure}

 From the time dependent wavefunction we construct the dipole 
 acceleration 
\begin{figure}
\pic{8}{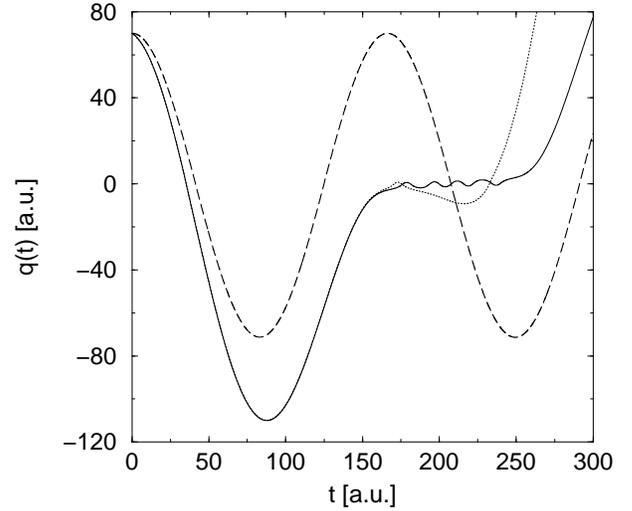}
\vspace*{0.1cm}
\caption[]{Examples for direct (dashed line), stranded (dotted line) 
and trapped (solid line) trajectories, see text.}
\label{trajek}
\end{figure}
\noindent
 \be{dipol}
d(t) = - \left < \Psi(t) \left | \frac{dV(x)}{dx} \right | \Psi(t) \right > 
\, ,
\ee
from which the harmonic spectrum 
\be{spec}
\sigma(\omega) = \int d(t)e^{i\omega t}dt
\ee
is obtained by Fourier transform. Typically $10^6$   trajectories are
necessary to converge $d(t)$ from \eq{dipol}. For comparison we have also determined   $d(t)$ 
quantum mechanically  (\fig{dipolpct}) using standard Fast Fourier Transform split
operator methods (FFT)  
to compute the wavefunction $\Psi(x,t)$. 

Figure \ref{spectrum} demonstrates that 
a plateau and a cutoff are visible in
the quantum (a),  and in the semiclassical (c) harmonic spectrum, 
but not in the classical (b) one. Since the semiclassical spectrum 
(b) and  the quantum spectrum (a) are very similar we  may conclude
that HHG can be described semiclassically.  Furthermore, the 
absence of 
the plateau in the classical spectrum (b) suggests that it is due to 
an interference effect of different types of classical trajectories 
contributing to the semiclassical result (c).  

Among the classical trajectories  from which the semiclassical 
dipole acceleration \eq{dipol} is constructed
we can distinguish trajectories which suffer a time delay when passing
the nucleus (i.e.\ $x\approx 0$) from 
the ``mainstream'' trajectories 
 which are not slowed down. Furthermore, among the time-delayed 
 trajectories we
can identify two groups. 

Trajectories of the first group (dotted line 
in \fig{trajek}) get  ``stranded'' on top of
the  barrier of the effective potential $V_{eff}(x) = V(x) -E_{0}x$.
 The second  group  is formed by trajectories which become temporally 
``trapped'',  (solid line in \fig{trajek}). 
The trapped trajectories are chaotic in the sense of an
extreme sensitivity to 
a change in initial 
\twocolumn[
\begin{figure}
\pic{16}{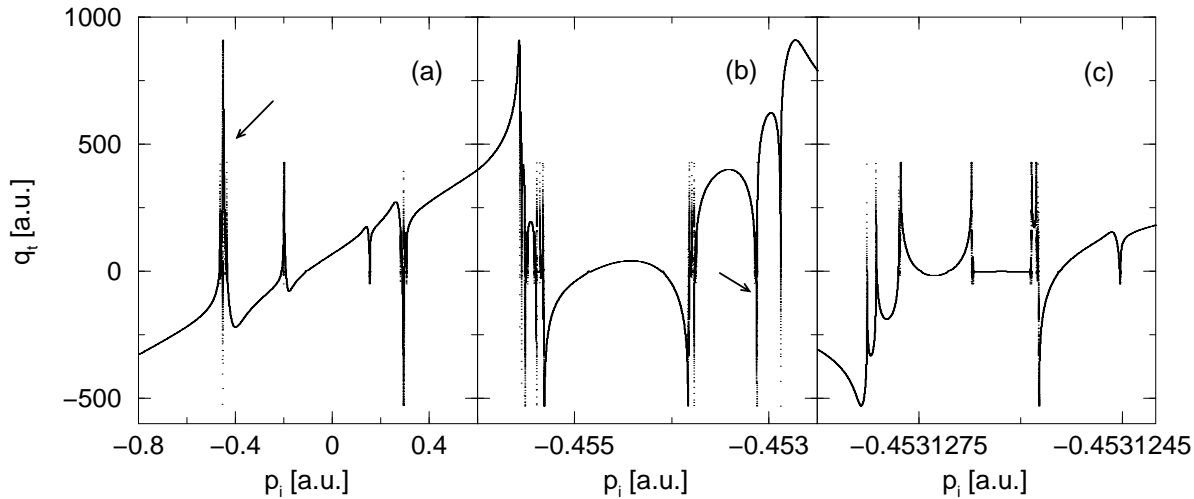}
\begin{minipage}{17.0cm}
\caption{Deflection function $q_t(p_i)$ for $t = 3T$ and $x_i = x_0 = 70 \, a.u.$ 
demonstrating the chaotic character of the trapped trajectories.
The arrows indicate the range of the next higher enlargement.} 
\label{deflect}
\end{minipage}
\end{figure}
]
\noindent
conditions. This is
clearly seen in the deflection function  $q_t(p_i)$ (\fig{deflect}) where
the final position $q_t$ of a trajectory at fixed time $t$ is 
plotted versus its initial momentum $p_i$ \cite{comment}.

One sees that in certain
intervals of $p_i$  small changes in $p_i$ lead to a completely
different $q_t$ with the result that the deflection function exhibits
a fractal structure. The fractal initial conditions (for a fixed final
$q_t$) belong to those  trajectories which are
trapped in the potential for a certain dwell time (the solid lines in \fig{trajek}). 

The time-delayed irregular orbits are responsible for 
the higher harmonics since their contributions
interfere with those from the mainstream trajectories. The interference manifests itself in 
a   dephasing in the dipole response $d(t)$ of \eq{dipol} after the 
first encounter with the nucleus 
 (roughly after the time $ t = T \equiv 2\pi/\omega$ for our initial conditions) 
 as can be seen on 
 \fig{dipolpct}b. At this time the peak
 at about $p_i\approx -.45 \, a.u.$, 
 emerges in the 
 deflection function, see \fig{deflect}. This corresponds to the 
 return of the nucleus in the case of an initially 
bound electron  as discussed, e.g., 
 in \cite{Cor93,Lew94}. The rich structure of this peak emerges 
 for longer times (see \fig{deflect})  necessary to resolve the
 fractal dynamics on a fine scale of the initial conditions $p_{i}$.
 The dephasing in $d(t)$ is clearly an interference phenomenon since
 it does not occur in the classical dipole response (\fig{dipolpct}a).

\begin{figure}
\pic{8}{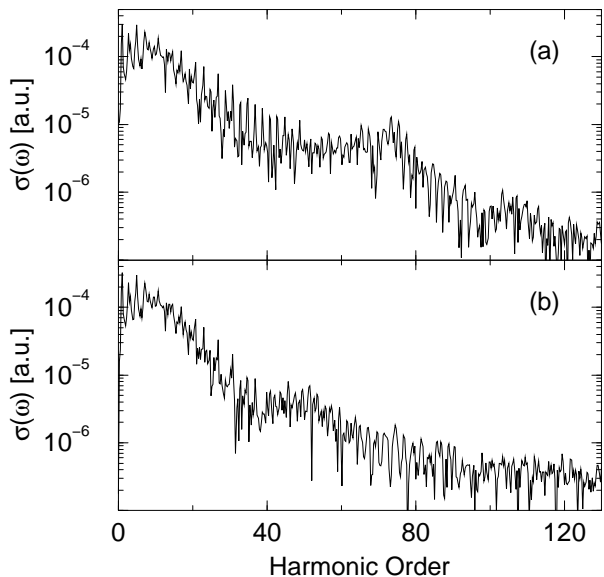}
\caption[]{Semiclassical HHG spectrum as in \fig{spectrum} but without 
trapped trajectories (a), and without time-delayed trajectories (b), 
see text.} \label{trapped}
\end{figure}

 Having  identified the orbits, or equivalently, the initial 
 conditions, which are responsible for the higher harmonics 
 we can artificially construct a harmonic spectrum without those
 contributions to double check that they are really responsible for 
 HHG. This has been done in the semiclassical spectrum of 
\fig{trapped}b where the
time-delayed trajectories (about 3\% of all initial conditions)  
have been discarded. Clearly, 
the plateau has disappeared and the spectrum is similar 
to the purely classical spectrum
with trajectories for all initial conditions included (\fig{spectrum}b). 
Discarding only the trapped 
trajectories (0.6\%) smears out 
the cutoff and leaves a reduced plateau for lower harmonics 
(\fig{trapped}a).
Hence, the quantitative semiclassical reproduction of the quantum 
HHG spectrum together with the absence of higher harmonics 
in the classical case (\fig{spectrum}) {\it and} in the semiclassical case
if irregular, time-delayed trajectories are discarded (\fig{trapped}) 
confirms our explanation of the origin of the higher harmonics.

To summarize we have shown that higher harmonic 
\noindent
\noindent
generation can be 
interpreted as a semiclassical interference effect between regular and 
time-delayed trajectories of the electron.  The time-delay is either 
due to a temporal trapping which generates chaotic dynamics or due to 
a stranding on top of the potential barrier.  Along with 
this time 
delay goes a characteristic difference in action compared to the 
undelayed mainstream orbits.  Analytical quasiclassical approximations 
of various kinds have been used to derive this phase difference which can explain 
the cutoff \cite{Cor93,Lew94}.  However, as demonstrated here, the 
full semiclassical expression is far more complicated since for the 
HHG spectrum  chaotic trajectories exhibiting a fractal 
deflection function are essential. The chaotic 
character of the irregular orbits allows them to have a relative
large effect in comparison to their weight among all initial 
conditions (of the 
order of 1\%) because their instability 
leads to a dramatic increase of their weight
$R_{\gamma}$ in \eq{monofak} in  the course of time. 
This increase makes an \newpage 
\noindent
accurate 
semiclassical computation
rather difficult. Remarkably, despite the chaotic 
dynamics of the trapped trajectories, one can obtain a converged 
semiclassical spectrum if a proper semiclassical propagator such as 
the Hermann-Kluk propagator is used which does not break down at the 
(abundantly occuring) caustics.  The resulting semiclassical harmonic 
spectrum agrees well with the quantum 
spectrum.

We would like to thank O.~Frank for the calculation of the quantum spectrum
and  C.~H.~Keitel  for helpful discussions.
Financial support from the  DFG under the 
Gerhard Hess-Programm and the SFB 276 is  gratefully acknowledged.

\end{document}